\documentclass[sigconf]{acmart}
\renewcommand\footnotetextcopyrightpermission[1]{} % removes footnote with conference information in first column
\AtBeginDocument{%
  }
\usepackage{algorithm} 
\usepackage{algpseudocode} 
\usepackage{multicol}
\usepackage{array}
\usepackage{pifont}
\newcommand{\xmark}{\ding{55}}

\begin{document}

\title[Never Trust the Manufacturer, Never Trust the Client]{Never Trust the Manufacturer, Never Trust the Client: A Novel
Method for Streaming STL Files for Secure Additive Manufacturing}

\author{Seyed Ali Ghazi Asgar}
\affiliation{%
  \institution{Department of Electrical and Computer Engineering, Texas A\&M University}
  \city{College Station}
  \state{TX}
  \country{USA}
}
\email{alighazi@tamu.edu}

\author{Narasimha Reddy}
\affiliation{%
  \institution{Department of Electrical and Computer Engineering, Texas A\&M University}
  \city{College Station}
  \state{TX}
  \country{USA}
}
\email{reddy@tamu.edu}

\author{Satish T.S. Bukkapatnam}
\affiliation{%
  \institution{Department of Industrial and Systems Engineering, Texas A\&M University}
  \city{College Station}
  \state{TX}
  \country{USA}
}
\email{satish@tamu.edu}

 \renewcommand{\shortauthors}{AAAA et al.}

\begin{abstract}
 While additive manufacturing has opened interesting avenues to reimagine manufacturing as a service (MaaS) platform, transmission of design files from client to manufacturer over networks opens up many cybersecurity challenges. Securing client’s intellectual property (IP) especially from cyber-attacks emerges as a major challenge. Earlier works introduced streaming, instead of sharing process plan (G-code) files, as a possible solution. However, executing client’s G-codes on manufacturer’s machines exposes them to potential malicious G-codes. This paper proposes a viable approach when the client and manufacturer do not trust each other and both the client and manufacturer want to preserve their IP of designs and manufacturing process respectively. The proposed approach is based on segmenting and streaming design (STL) files and employing a novel machine-specific STL to G-code translator at the manufacturer’s site in real-time for printing. This approach secures design and manufacturing process IPs as demonstrated in a real-world implementation.
\end{abstract}





\maketitle

\section{Introduction}

        Additive manufacturing(AM) or so-called 3D printing is a method to  make physical objects by stacking materials layer by layer using a 3D Computer-Aided Design (CAD) model. Initially, AM was used for rapid prototyping; however,  nowadays it is used for finished product manufacturing as well. AM also eliminates extensive processes required by traditional machining methods \cite{gibson2021additive}. This layer-by-layer approach makes it possible for engineers to design and manufacture complex structures with reasonable geometric quality and precision \cite{tofail2018additive,debroy2018additive}. 
    
        Recent advances in cyber-physical systems, computer networks, and smart manufacturing technologies are driving the emergence of manufacturing as a service (MaaS) platforms \cite{tolio2023platform}. To clarify this new topic we will provide a simple example. Assume that there is a car manufacturing company called \textit{GoodCarMakers} and its manufacturing facility is located in Europe. There is also a car owner in North America who has been using the same car for the last 20 years manufactured by the \textit{GoodCarMakers}. Now, a gear in the car window lifter is broken and the car owner asks the \textit{GoodCarMakers}'s customer service to repair the car. \textit{GoodCarMakers} could have kept this gear in the inventory for the last 20 years. However, this solution costs money, space, labor, and international transportation. The other option is that since the car company has the digital design file of the gear, they can use 3D printers to manufacture the gear whenever it is necessary at the point of need\cite{iquebal2018towards,Gupta_Supply_Chain}. The company can use their own fabrication facility in Europe to make the part and transport it to North America or they just can share the file with their customer service at North America to eliminate transportation costs. It is also possible for the \textit{GoodCarMakers} to send their file to another third-party company in North America which is an expert in additive manufacturing and ask them to produce the gear for them.

        The concomitant growth of additive manufacturing (AM) into an estimated \$2 trillion market by 2030 \cite{prashar2023additive}  further helps manufacturers to produce on-demand items in distributed sites with shortened supply chains \cite{zanardini2016additive}. Although these emerging MaaS platforms can reduce costs and time, they are increasingly vulnerable to security risks\cite{bauer2017movement}. 
        
      As the digital version of the product is shared across the supply chain, attackers may steal the computer-aided design (CAD) files, disregarding copyright, and intellectual property (IP) rights. Once compromised, an attacker could sabotage the digital thread by altering the properties of the part in the CAD file, or editing the process plan (G-Code) files. For instance, a Trojan attack on 3D printers\cite{Trojan}  can compromise the quality of a 3D printed component, reducing its tensile strength by 50\%. The strength of a 3D printed component can be compromised extensively by changing the orientation of printing leading to failure of the test component \cite{zeltmann2016manufacturing}. 
      Additionally, an attacker can convert stolen CAD files into Standard Triangle Language (STL) files and produce counterfeit items without the need for reverse engineering. While these counterfeit items may appear identical to the original components, they often lack the material integrity and durability of the genuine products. If these counterfeit parts enter the supply chain, they could cause serious harm, leading to equipment damage and personal injuries \cite{kurfess2014rethinking} . Attackers could also sell digital files to other countries or industrial competitors. Reports also suggest that 60\% of these attackers are former employees \cite{warren2015modern}.
      
\begin{figure}
    \centering
    \includegraphics[width=1\linewidth]{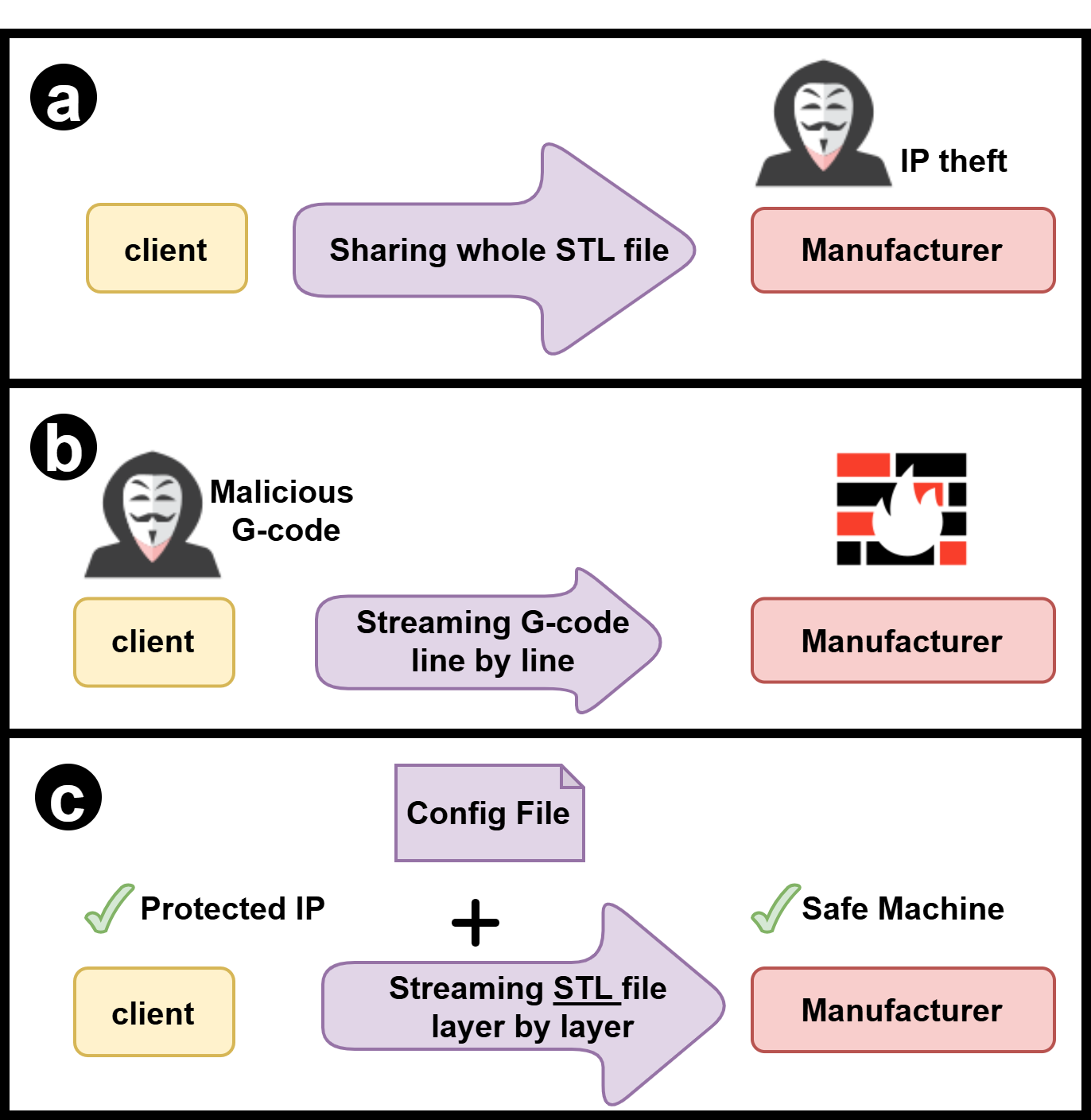}
    \caption{(a) The shares the whole file with the manufacturer. (b) streaming the G-code file line by line. (c) streaming the STL file layer.}
    \label{fig:stlstream}
\end{figure}

      IP thefts cost companies more than \$200 billion each year \cite{warren2015modern}[10]. Therefore, it is necessary to consider a secure method for managing the digital files in the manufacturing supply chain. 
     A growing solution to protect and authenticate design files from sabotage, counterfeiting and espionage is to embed codes within the design files \cite{chen2021embedded,tiwari2021protection,mahesh2020surveyreddy}. Once the component is manufactured, these concealed embedded codes can be extracted to validate the integrity and quality of the manufacturing process and product. However, any authentication method is reactive, as it does not mitigate the risk of IP theft during the transmission of the design and process plan files across the digital thread in a platform. A major IP risk is noted to arise when a client shares the entire design (“STL”) file with a manufacturer (see Figure \ref{fig:stlstream}(a)), potentially exposing it to theft or unauthorized replication \cite{mahesh2020surveyreddy}[13]. Another method to protect design files from malicious modifications is by taking advantage of the blockchain technology. Researchers in \cite{blockchain} proposed a dedicated framework between design and 3D printing companies for the management of the design files against IP thefts. 
     
To address this challenge, one method for transferring digital files from client to the manufacturer is handled through streaming. Previous works\cite{baumann2017model, tiwari2020cybersecurity}  suggested streaming the G-code file to AM machine to limit the manufacturer’s access to the original design file (Figure \ref{fig:stlstream}(b)). In this way, every single line of a G-code file is streamed separately, similar to video streaming services. 
Although this method enhances IP protection, it also introduces the risk of malicious clients sending compromised G-code files, which could lead to significant financial losses, safety hazards, and physical damage on the manufacturer side as G-code commands directly control the AM machine. 
As a result, clients seek a way to stream their files without disclosing the entire object design, while manufacturers prefer not to allow external G-code execution on their machines to mitigate the risks associated with untrusted G-code. In addition, the streamed G-code may be incompatible with the AM machine in terms of meeting the motion trajectory and machine limitations, or worse, create anomalous settings or trajectories that damage the machine. 
In this work we propose an approach to address security concerns across a MaaS platform, including of the client (designer) and the manufacturer. The contributions of this effort are as follows:
\begin {itemize}
    \item Propose an approach for protecting design IP of a client and the process IP of a manufacturer at the same time (when client and manufacturer may not trust each other)
    \item A novel approach for segmenting a design file into STL file segments for streaming and converting streamed STL data into the manufacturer’s machine specific G-codes.
    \item Evaluating the proposed method with parts that are printed with a specific angle and orientation.
    \item Developing techniques to segment STL file into horizontal pieces and generate custom support for the part.
    \item A real-world implementation of the proposed method on a fused filament fabrication 3D printer to demonstrate the method.
\end{itemize}
The rest of the paper is organized as follows. In Section \ref{sec:overview_terms} we will provide the general terms used in additive manufacturing area. We will talk about our threat model in  section \ref{sec:threat_model}. Section \ref{sec:research_approach} highlights the research approach, Section \ref{sec:Implementation} presents the implementation details and the results, and Section \ref{sec:conclusion} contains the conclusions.

% \subsection{Template Parameters}

% In addition to specifying the {\itshape template style} to be used in
% formatting your work, there are a number of {\itshape template parameters}
% which modify some part of the applied template style. A complete list
% of these parameters can be found in the {\itshape \LaTeX\ User's Guide.}

% Frequently-used parameters, or combinations of parameters, include:
% \begin{itemize}
% \item {\texttt{anonymous,review}}: Suitable for a ``double-anonymous''
%   conference submission. Anonymizes the work and includes line
%   numbers. Use with the \texttt{\string\acmSubmissionID} command to print the
%   submission's unique ID on each page of the work.
% \item{\texttt{authorversion}}: Produces a version of the work suitable
%   for posting by the author.
% \item{\texttt{screen}}: Produces colored hyperlinks.
% \end{itemize}

% This document uses the following string as the first command in the
% source file:
% \begin{verbatim}
% \documentclass[sigconf]{acmart}
% \end{verbatim}

\section{Overview of Technical Terms in Additive Manufacturing}
\label{sec:overview_terms}
\begin{figure*}[t]
    \centering
    \includegraphics[width=1\linewidth]{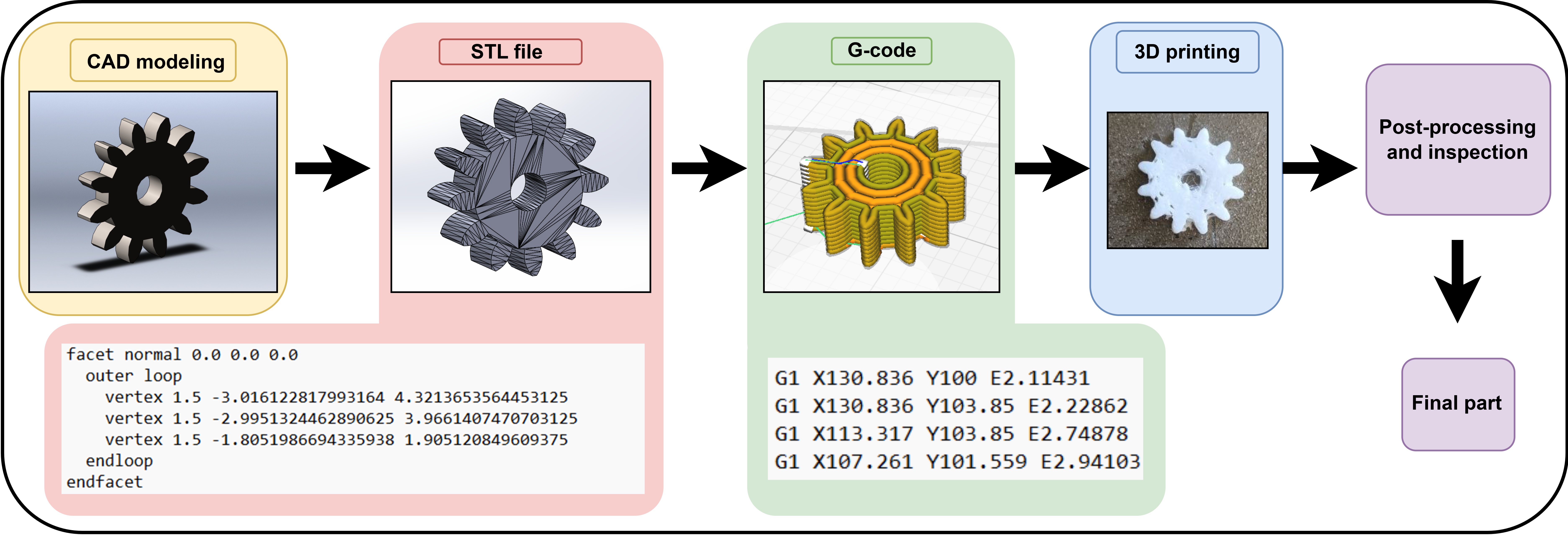}
    \caption{ The workflow of an additive manufacturing chain. In the first step, the file is designed in a CAD application. Then, it is converted into an STL file, and this STL file is fed into a slicer application where it is converted into G-code format. Finally, the G-code file is sent to the machine for manufacturing the part.}
    \label{fig:process}
\end{figure*}

In this section, we will go over each step in the workflow of AM chain (shown in Figure \ref{fig:process}) and explain their purpose.

\subsection{Computer-Aided Design(CAD) Applications}
In the initial stage, designers typically use 3D CAD software to transform their ideas and requirements into a 3D model. Each CAD software saves these designs in a unique file format. In this work we used SOLIDWORKS\textregistered \  for designing different parts, and saving them into STL format.

\subsection{Standard Triangle Language(STL) File}
The STL file format is one of the most commonly used formats in additive manufacturing. As implied by its name (Standard Triangle Language), it converts the original 3D design into a network of tiny triangular meshes, each with defined geometries and coordinates. These triangles, as illustrated in Figure \ref{fig:process}, fit together to approximate the shape of the original design. 
\subsection{Slicer Applications and G-Code Format}
STL files cannot be fed directly to an AM machine. Therefore, an intermediate application, called a slicer, receives the STL file as input and converts it to the G-code format based on several configurations. As shown in Table \ref{table:gcode_table}, G-code commands control various parameters of an AM machine, from moving the nozzle, setting the temperature of the hot end and restarting the machine to selecting files on the SD card, updating the firmware, etc. Essentially, every interaction is managed through G-codes. A G-code file is an aggregation of different G-code commands that when combined together will print the desired file for a user.

Slicer applications are responsible for converting an STL file into G-code. In the slicer, the user can also specify the desired configurations. Some of these configurations (see Table \ref{table:config_stl}) are related to the manufacturing machine, and some of them are related to the designer's choice. For instance, a designer does not need to know fan speed or bed temperature. On the other hand, designer should specify the amount of infill as it is directly related to the strength of the final product. In addition, layer height, which specifies the resolution of the printed part, is also another designer's choice for manufacturing finer components.     

\begin{table}[h]
\centering

% \captionsetup{justification=centering}
\caption{Example of G-Code that are widely used in Marline firmware for 3D-printers }
\label{table:gcode_table}

  \begin{tabular}{
  >{\centering\arraybackslash} m{1cm} 
  >{\centering\arraybackslash} m{2cm} 
  >{\centering\arraybackslash} m{4cm} }
  
    \toprule
 Command & Example & Description \\
    \midrule
G0 & G1 X10.2 & Move to the location 10.2mm on X axis  without extrusion \\

\hline
 G1 & G1 X10.2 Y14.3 & Move to the location 10.2mm on X axis and 14.3mm on the Y axis with extrusion on\\

\hline
 G12 & G12 & Clean the nozzle\\

\hline
 
 G28 & G28 X Y & Home X and Y axis \\

\hline
 G92 & G28 X0 Y0 & Specify the current nozzle x and y location to 0 and 0 \\
 \hline
 M23 & M23 gear.gcode & Select the file on sd card to print \\
 \hline
M106 & M106 S100 & Set the fan speed to 100 \\
\hline
M104 & M104 S250 & Set the the hotend temperature to 250 degree \\
\hline
M997 & M997 & Update the firmware from SD card \\
\hline
M999 & M999 & Restart the machine \\
% \hline
 \bottomrule

 \end{tabular}

\end{table}

% \end{multicols}

% \textcolor{red}{In current practice, an STL file is transferred to a manufacturer who then uses a slicer to convert that into machine-specific G-code for printing a part. This leads to potential IP theft, counterfeiting and other issues.}
In current practice, an STL file is transferred to a manufacturer who then uses a slicer to convert that into machine-specific G-code for printing a part. This leads to potential IP theft, counterfeiting and other issues.

\section{Threat Model}
\label{sec:threat_model}

As noted, protecting IP from malicious attackers who might steal or alter design and process plan files across the digital thread is a major security threat \cite{baumann2017model, tiwari2020cybersecurity} . An earlier work introduced streaming the G-code files directly to the AM machine from the client, similar to how video streaming software revolutionized the entertainment industry (see Figure \ref{fig:stlstream} (b)). The primary concern with this method is that the manufacturer has to trust client-generated G-codes. Second, the client may not know all the manufacturing secrets of producing parts on a given machine and the manufacturer would like to retain this know-how without disclosing it to the client to generate appropriate G-codes.

If the client is allowed to stream the G-code directly to the machine, a G-code file infected with malware, or a malicious client could intentionally alter the G-code to attack the AM machine and cause interruptions. For example, the nozzle temperature can be increased beyond safe operating environment.
Therefore, the main objective of this work is not only to protect the clients from IP theft, but also to protect the manufacturer's machines from executing untrusted G-codes.  Effectively, the threat model considered in this paper is as follows: a client/designer would like to protect IP from a potential IP theft while a manufacturer would like to protect their AM machines from malicious/untrusted G-code and protect the IP of the manufacturing process.

        To address this issue, we propose a new STL generation and streaming method instead of G-code streaming. To do so, we first convert a CAD file into a single STL file and then we split the STL file into multiple STL sections, corresponding to different layers, and each such segmented STL file is sequentially streamed to the manufacturer. Once received on the manufacturer side, the manufacturer is responsible for converting this STL file to the G-code file, which eliminates the danger of direct execution of untrusted G-code commands.

\begin{figure*}[t]
    \centering
    \includegraphics[width=1\linewidth]{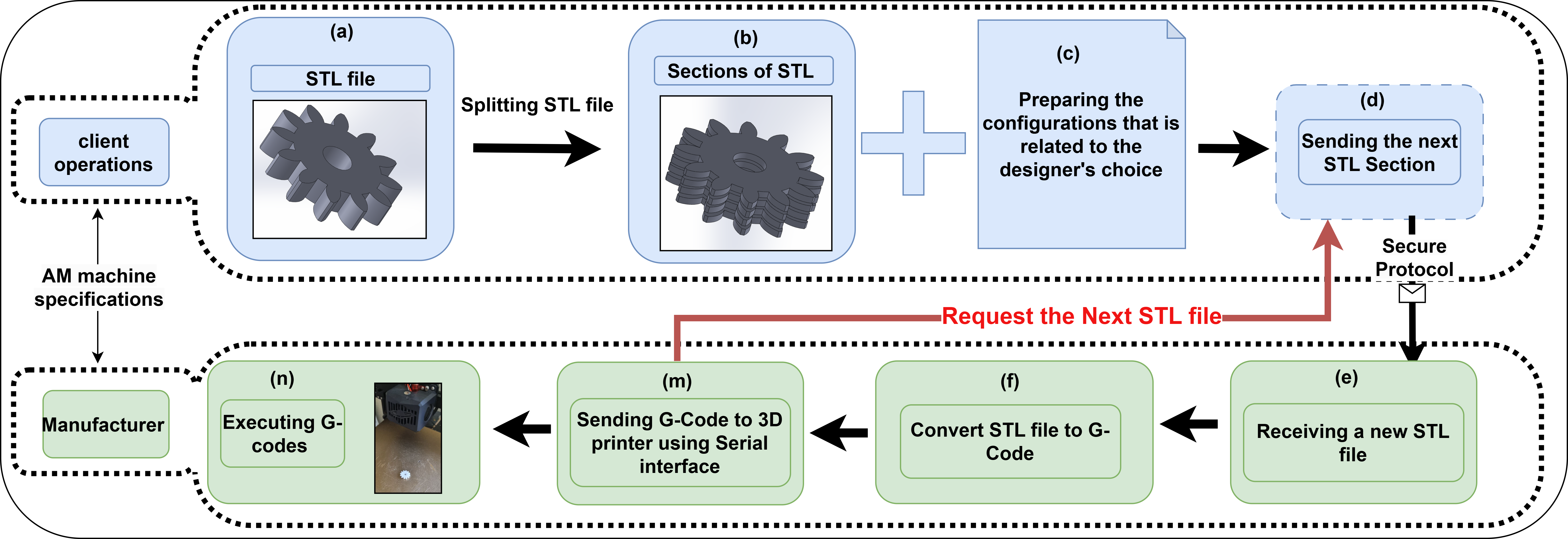}
    \caption{ First, client asks the manufacturer for the machine specifications. Then, (a) the client designs the file and converts its desgin to a STL file. Then, (b) the client slices the design file using our custom Python scripts. After that, (c) the client creates the configuration. Finally, (d) the first section of STL file is sent to the manufacturer through a secure channel. (e) Once the STL file is received , the manufacturer uses the slicer application along with the configuration files to produce the G-code file (f). The G-code file is further processed , and the data is sent to the printer via serial communication (m). Once the first layer is printed, the manufacturer requests the next layer’s STL file, and the cycle repeats.}
    \label{fig:structure}
\end{figure*}

\section{Research Approach}
\label{sec:research_approach}

\begin{table}[b]
\centering

% \captionsetup{justification=centering}
\caption{Some of the configurations that are used in slicer applications \cite{githubGitHubSlic3rSlic3r}}
\label{table:config_stl}

  \begin{tabular}{
  >{\centering\arraybackslash} m{2.8cm} 
  >{\centering\arraybackslash} m{2.2cm} 
  >{\centering\arraybackslash} m{2.2cm} }
  
    \toprule
 Configuration Name &  Machine choice & Design choice  \\
    \midrule
Bed temperature & \checkmark & \xmark   \\
Bridge fan speed & \checkmark &   \xmark \\
Disable fan    & \checkmark & \xmark   \\
Filament diameter & \checkmark &   \xmark \\
Temperature & \checkmark &  \xmark  \\

Nozzle diameter & \checkmark &  \xmark  \\
Fill angle & \xmark &  \checkmark  \\
Fill gaps &  \xmark &  \checkmark  \\
Fill density & \xmark &  \checkmark  \\
Fill pattern & \xmark &  \checkmark  \\
Layer height  & \checkmark &  \checkmark  \\
Infill every layers & \xmark &  \checkmark  \\
\bottomrule

 \end{tabular}

\end{table}

An STL file is not organized in a bottom-to-top layering pattern as in a G-code file. This results in a lack of dependency recognition between different sections within an STL file. Consequently, the main challenge is to find an appropriate solution for streaming the STL file. To solve this issue, we conceived a novel method to split a single STL file into horizontally sliced STL parts and then stream the segmented STL files (see Algorithm \ref{alg:client} ) at the client’s end and to convert the streamed STL files in real-time to machine specific G-codes (Algorithm \ref{alg:machine}).
Algorithm 1 involves first orienting the file and then dividing the original STL file into horizontally stacked sections and then exporting each section as an individual STL part. In the initial step, horizontal planes are established along the surface that will adhere to the printer's bed. These planes are then duplicated at specific intervals. For example, if a layer height h = 0.3 mm is chosen for 3D printing, the spacing between planes should also be set to be approximately 0.3 mm. 

Layer height is determined by design needs and manufacturing machine capabilities, and hence this is determined as part of the config file in the initial handshake between the client and the manufacturer. The layer height is important for both the manufacturer and client. The client needs to set up the layer height based on the resolution needed; however, the client must be aware of the AM machine’s limitations. Therefore, before sectioning a STL file, the designer and manufacturer agree on a valid range of this variable.

We developed a custom STL slicer that gets an STL file and layer height as input and then creates \textit{k} STL files with equal height as the output.  For instance, if the model's height is 3 mm and we aim for a 0.3 mm layer height resolution, this process will yield ten separate parts.
Specific printing configurations such as layer height, infill percentage, and other relevant settings must be defined by the designer. These parameters are saved in a configuration file, which is then sent to the manufacturer to be used during the slicing process during manufacturing. 

Once the STL file is received from the client, the manufacturer executes Algorithm \ref{alg:machine} which first requires deleting any previously stored G-code and STL files. Keeping multiple files could enable the manufacturer to combine them and cause IP theft. This requirement applies to all streaming applications; if the receiver stores all incoming data, the security of the streaming process is fundamentally undermined. At the manufacturing site a custom slier is used to convert the STL file into a G-code file (e.g., 'body1.stl' to 'body1.gcode'). During this process, the appropriate configurations are provided to the slicer tool that satisfies both the designer's and the manufacturer's criteria.
        After generating the G-code file, additional commands at the beginning and end of the file are removed to optimize the printing sequence. Typically, G-code files from standard converters contain initial commands for setting the printer temperature and calibrating the home position at the start of the file. However, these commands are only necessary for the initial file and thus are omitted in subsequent files. Similarly, redundant commands at the end of the G-code file, such as those for turning off the nozzle and bed and moving the actuator to the (0,0) location, are only required for the last file when the printing process is completed. Hence, these commands are removed from intermediate files to prevent unnecessary actions. We developed custom scripts to remove these redundant commands from the beginning and end of the intermediate files to automate the process. 

It is worth noting that each manufacturing machine has its own specifications; therefore, appropriate configurations must be used for a specific machine. Hence, options such as nozzle temperature, filament diameter, and printing speed must be set properly in the application on the manufacturer’s side. A manufacturer can change the settings based on the machine’s conditions for the first time and keep it unchanged for the rest of the manufacturing requests. 

        In this work, the manufacturing machine is connected to the computer via serial communication. Serial communication is commonly used in 3D printing to interface with printers and transmit G-code commands. We utilized the PySerial library to establish communication with the printer's serial port and send G-code commands to the AM machine. 
        Once the printing of the first layer is complete, the manufacturer requests the next STL file. The key difference for the new iteration is that during the slicing process, the Z-offset value, i.e., the height at which printing begins, must be properly adjusted. Since the previous STL file has already been printed, the new file should not start at Z=0 mm but rather at Z=h (the layer height with a suitable allowance for shrinkage and material overflow), positioned above the previous layer. Consequently, for each subsequent STL file, the Z-offset is incremented by +h, ensuring that each new layer is printed on top of the previous layers.

\begin{algorithm}[b]
	\caption{Client's logic} 
        \label{alg:client}

        \begin{algorithmic}[1]
            \State $Specs$ = GetMachineSpecifications($Manufacturer$)
            \State $Config$ = GenerateConfigFile($Specs$)

            \State $STL$ = OrientTheSTLPart($STL$)
            \State $STL$ = GenerateSupportScript($STL$)
            \State $STLArray$ = ConvertSTLtoSections($STL$)
            
            \State SendConfigFile($Config$,$Manufacturer$)
            \For {$n=1,2,\ldots$}
                \State Stream(STLArray[$n$], $Manufacturer$) 
                \State Wait() //until manufacturer requests the next file
            \EndFor
	\end{algorithmic} 
\end{algorithm}

\begin{algorithm}[t]

	\caption{Manufacturer's logic} 
        \label{alg:machine}
    
	\begin{algorithmic}[1]
                \State SendMachineSpecifications($Client$)
                \State $config$=GetConfigFile($Client$)
                \State $STLArray[0]$=GetSTLFile($Client$,0)
                \State \textit{Z\_offset} = 0
		\For {$n=0,1,2,3,\ldots$}
                \State $STLArray[n+1]$=Thread(GetSTLFile($Client$,n+1)) 
                \State RemovePrintedfiles()
                \State $Gcode$ = SliceSTLfileToGCODE($STLArray[n]$, $config$)
                \State \textit{Z\_offset} += \textit{config.layer\_height}
                \State \textit{config.Z\_offset} =  \textit{Z\_offset}
                
                \State $Gcode$ = RemoveRedundantCommands($Gcode$)
                \For {$k=1,2,\ldots $}
                    \State SendCommandOverSerial($Gcode$, $line$ = $k$)
                \EndFor
            \EndFor

	\end{algorithmic} 
\end{algorithm}

\begin{figure}[t]
    \centering
    \includegraphics[width=1\linewidth]{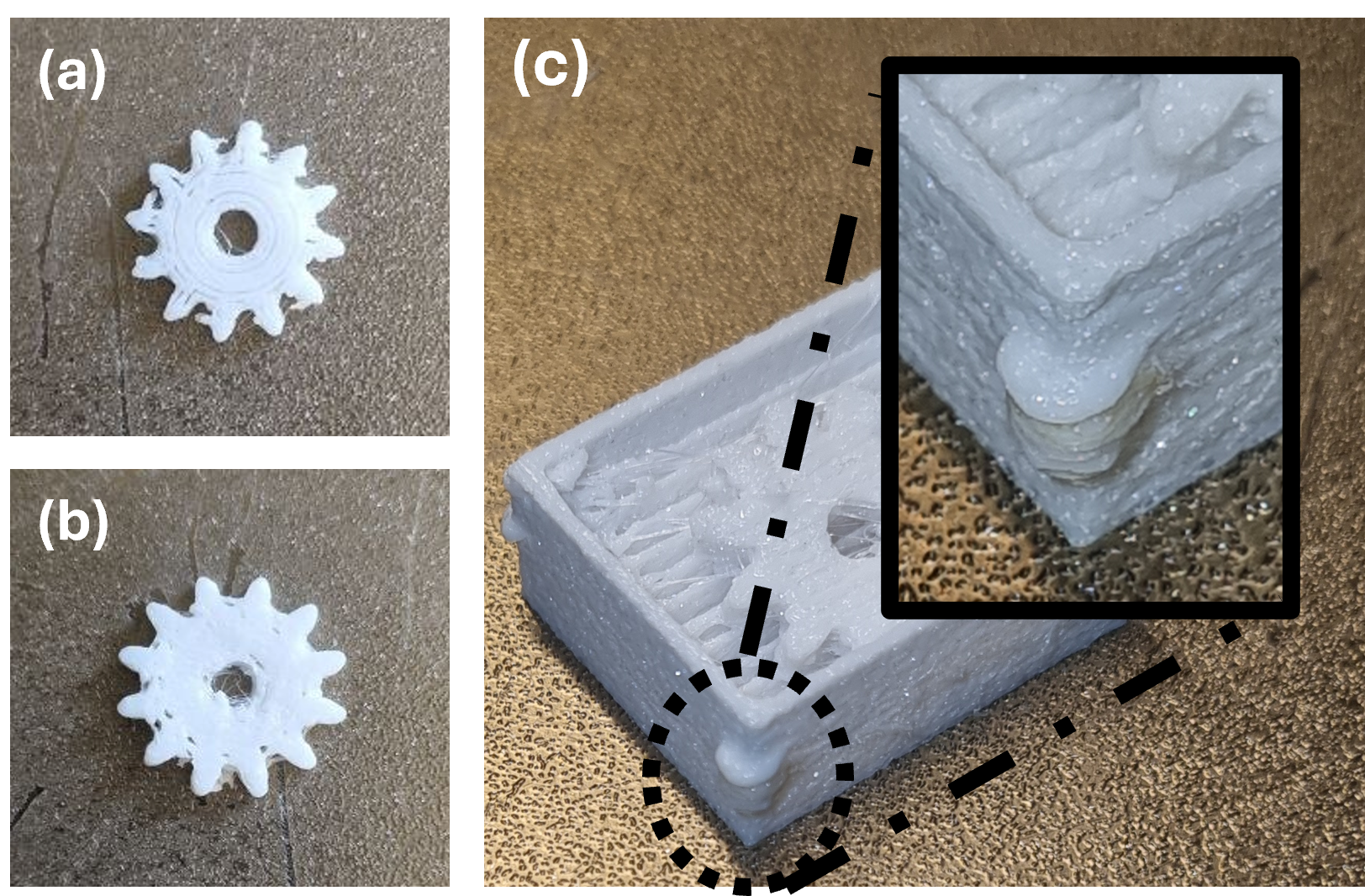}
    \caption{(a) Streaming method and (b) normal print. (c) The effect of latency in streaming files. When the nozzle stops and waits for the next chunk of data, extra filament flows from the nozzle at the stopping point.}
    \label{fig:simple_latency}
\end{figure}

\subsection{Security Advantages of the Proposed Approach}
One benefit of the proposed approach is that the streaming process can be done through multiple distributed servers to the manufacturing side \cite{ tiwari2020cybersecurity}. This reduces the risk of communications-related interruptions compared to when the client's sole server is attacked. 
In addition, since the client is forbidden from directly executing G-code commands on the manufacturer's machine, the MaaS platform is protected against cyber-attacks such as command injection and Man in The Middle (MITM),  which could directly affect the AM machine. On the other hand, in previous work \cite{ tiwari2020cybersecurity}, if an attacker could intercept the connection between the client and manufacturer and change the normal G-code commands to malicious commands, it could damage the AM machine as the G-code was streamed directly. Furthermore, since we are minimizing human intervention during printing and limit manufacturer's access to the design file, we enhance the security of the design files against IP theft and counterfeit production. 
In this work, we assume that the connection between the manufacturer's computer and AM is secure, i.e., the machine operator would not place a sniffing device between AM machine and streaming computer.  In addition, we expect that malicious workers do not have access to the manufacturer's computer memory. Otherwise, it is possible to save all the streamed files together and reconstruct the original file. This assumption amounts to having the controller of the AM machine with streaming capabilities, i.e., a single device controls the printing trajectory as well as data streaming.

\section{Implementation}
\label{sec:Implementation}

% \subsection{Streaming an archetype Gear} 
        For a basic archetype, we examined the streaming of a simple gear to the manufacturer for 3D printing. We used an Elegoo Neptune 3 printer with Poly-lactic Acid (PLA) filament. We set the layer height as 0.3 mm with 100\% infill density and the printing temperature was set to ~250°C. The manufactured gear in both the normal scenario (printing with the whole G-code file) and streaming case are shown in Figure \ref{fig:simple_latency}. As shown in the figure, there is a marginal difference in the streamed part compared to the normal part ($\pm$0.3 mm), demonstrating the effectiveness of the proposed method. 
        
        % \textcolor{red}{Can we quantify any differences or how close the two methods are?}

\subsection{Streaming Complex Parts with Supports and Special Orientation} 

\begin{figure}[t]
    \centering
    \includegraphics[width=1\linewidth]{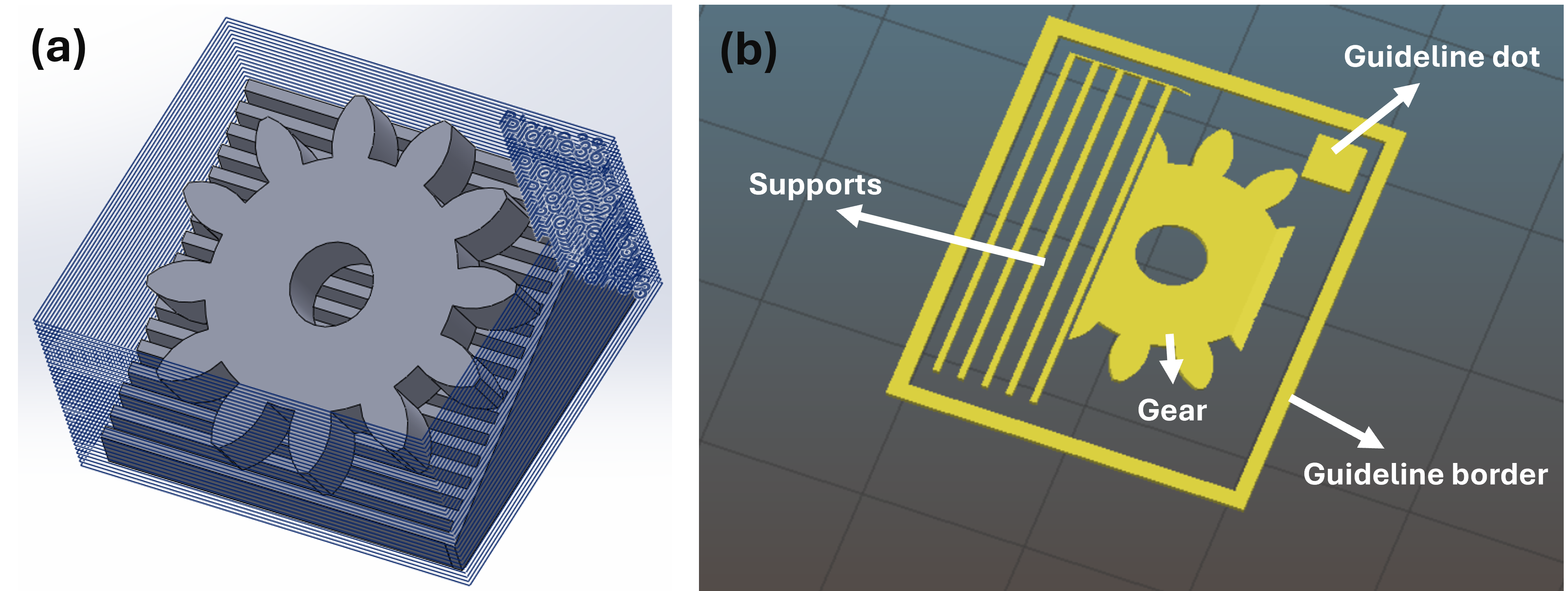}
    \caption{(a) Sample component with special orientation and support. (b) Special guideline dot and guideline border for solving the dislocation problem. This figure is also an example of a single STL file.}
    \label{fig:guideline}
\end{figure}

\begin{figure}[b]
    \centering
    \includegraphics[width=1\linewidth]{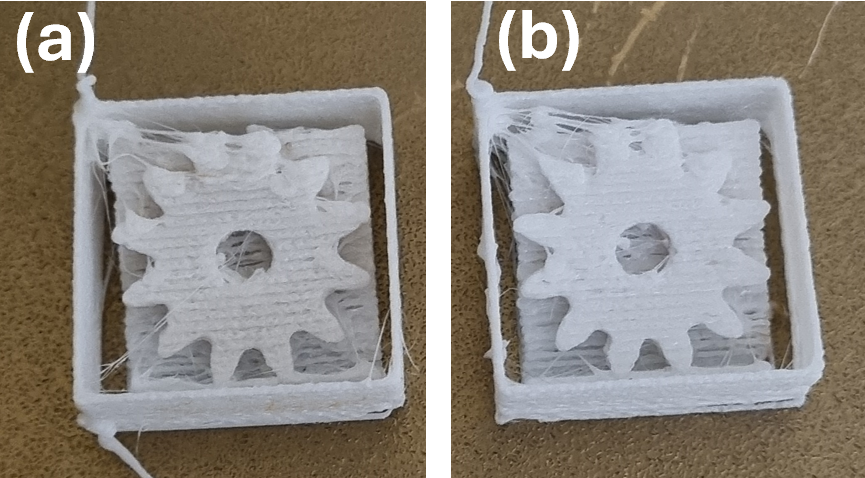}
    \caption{Manufactured object from (a) streamed data with special guideline dot and guideline border for solving the dislocation problem. (b) normal method for manufacturing the part (support is generated using our python script).}
    \label{fig:streamvsnormal3}
\end{figure}

        Although the simple gear case was successfully manufactured, we should also consider more complex cases that involve special orientations or require external support. Typically, slicer applications automatically generate support for the necessary areas. However, in our case, support cannot be generated during the slicing process because this step is handled by the manufacturer. The manufacturer only has access to a single layer of the file at a time, not the complete file. As a result, the manufacturer has no prior information about which surfaces are printed on top of which previous layers or whether supports are needed for a specific location. To address this problem, we wrote a  script to generate supports for a STL file (see Figure \ref{fig:guideline}(a)). With this method, each STL file includes the support components, eliminating the need for the slicer application to generate or have any knowledge about the supports. 
        %We automated this process by using a custom Python scripts that receives a file and generate the needed support at the bottom of the STL file. 
        In our approach, the needed supports are generated at the client as part of the STL file. 
       Another issue that we noticed with the slicer application on the manufacturer’s side is that different layers could be imported at varying locations on the XY plane. For example, the first layer might be printed at the location (0,0) based on its center of surface, while the second layer could shift to a different position, such as (0.1, 0.2), due to the differences in its surface or borders. 
To address this, we introduce the idea of a guideline border which is basically a fixed rectangle around the printed object (See Figure \ref{fig:guideline}(b)). This ensures that each layer has a consistent outer boundary which, based on our experiments, keeps the imported file aligned in the same position for every layer. Additionally, we added a small rectangular dot for debugging purposes. This dot helps confirm that the file is not rotated and that the rectangle is printed consistently at the same location for each layer. 
      The result of printing this complex design is shown in Figure \ref{fig:streamvsnormal3}. Here, the final part is manufactured with special orientation and support without any issues.

\begin{figure*}[!t]
    \centering
    \includegraphics[width=1\linewidth]{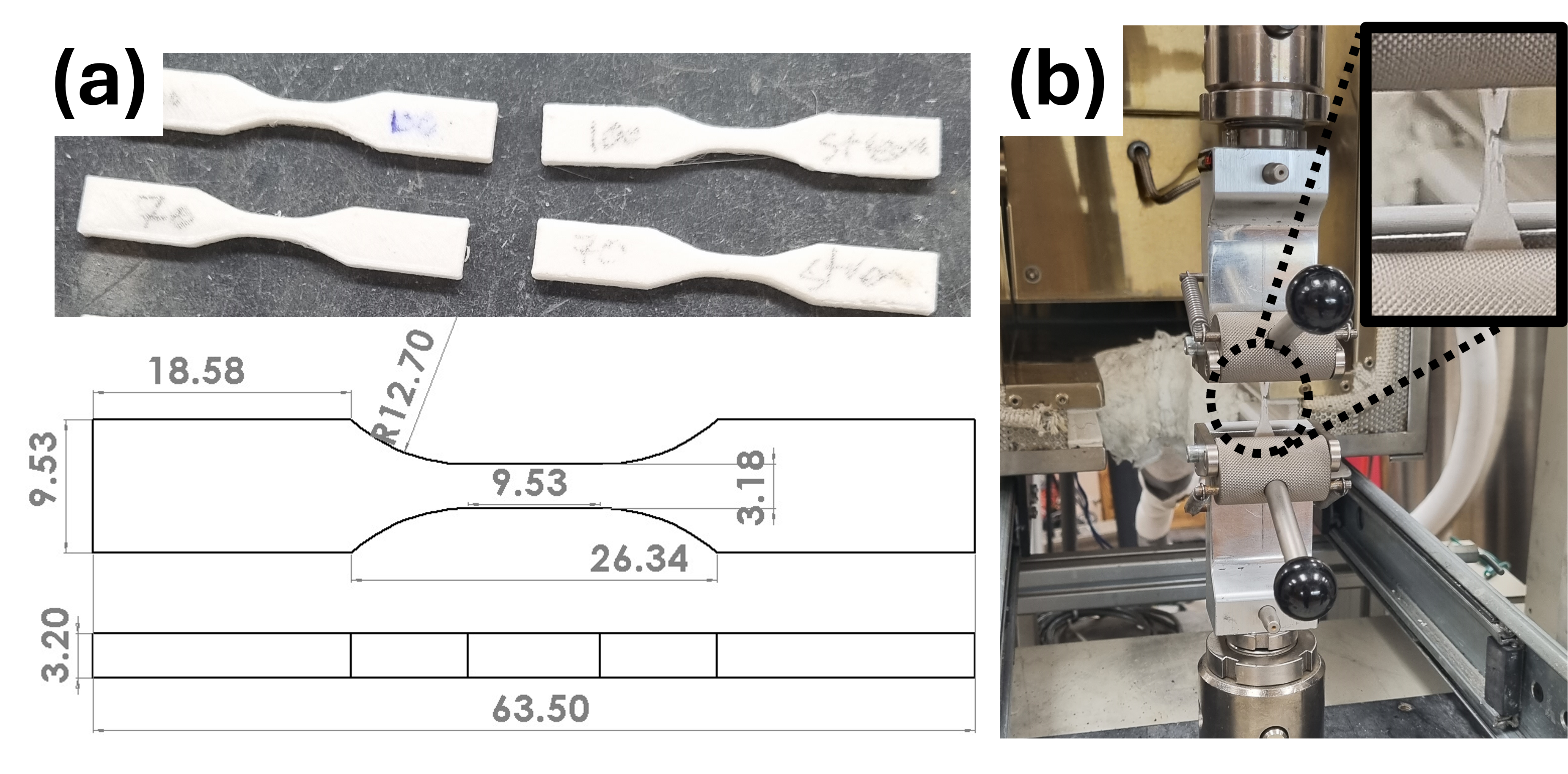}
    \caption{(a) Dimension of the specimen (D638 V) and printed parts with 100\% and 70\% infill in normal and streaming methods. (b) Tensile test performed on the specimen. }
    \label{fig:labtest}
\end{figure*}

\subsection{Streaming Related Latency
} 
        Latency issues are always present in streaming applications. Since we are sending the commands in real time, if printing is finished before the next layer is ready, the printer's nozzle stops at the last point. This stoppage may cause extra leftover filament from the nozzle's end to flow unnecessarily on top of the previous layers. This problem creates extra bumps in the printed file and might deform the object (see Figure \ref{fig:simple_latency}(c)). We noticed that during the printing phase, as the printer waits for the next command to arrive, extra filament melts at this location, causing this unwanted issue. To solve this problem, we used the buffering technique. In the beginning, we request the first two layers at the same time and wait until we fully receive both of them and both are converted into G-code. Then, while the second layer is being printed, we request the next layer by calling a separate thread for the third layer. Hence, while the nth layer is being printed, we are receiving and converting the (n+1)th STL file to G-code.

\subsection{Results}
\begin{figure}[!b]
    \centering
    \includegraphics[width=1\linewidth]{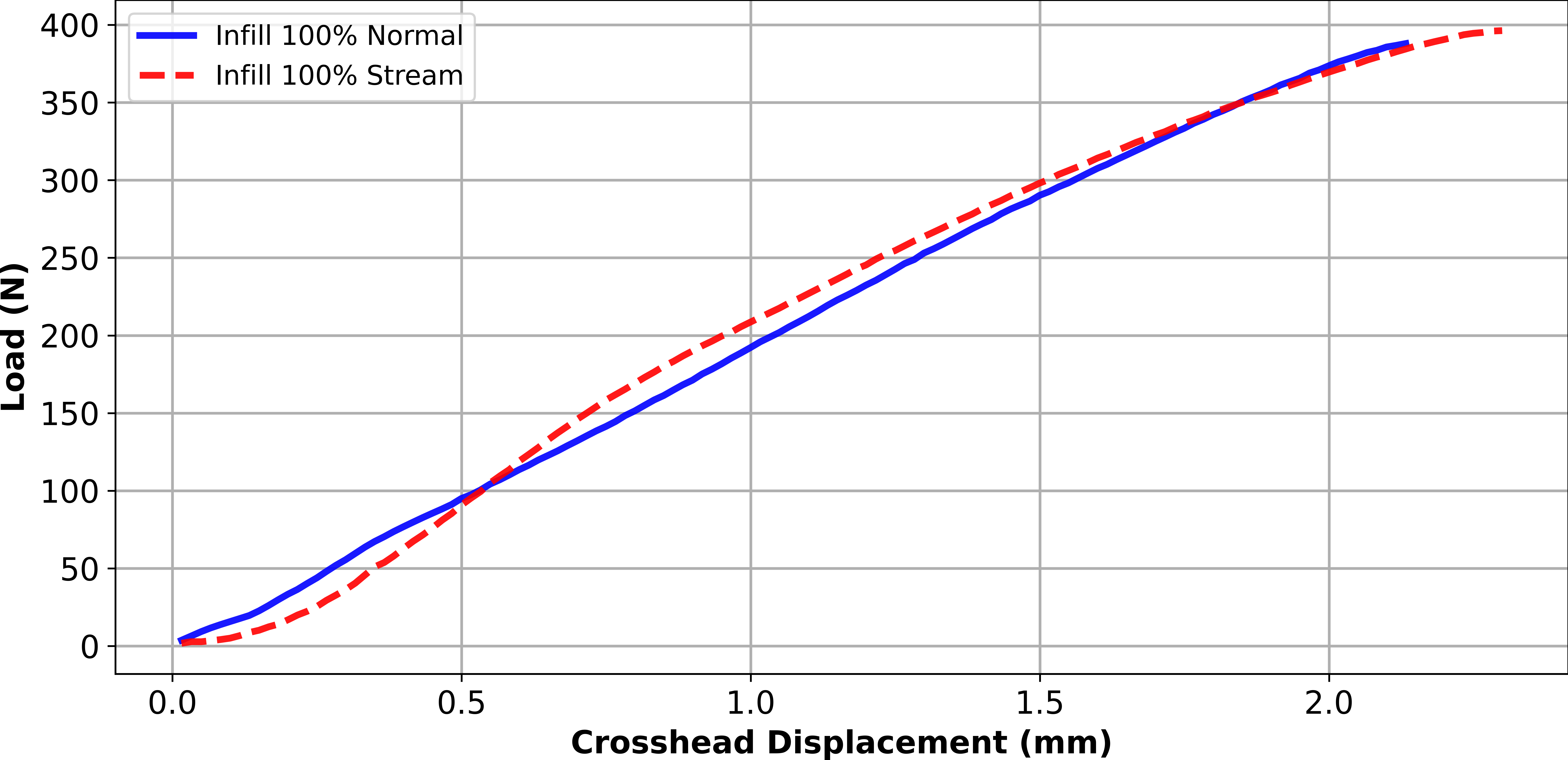}
    \caption{Force/displacement curves for tensile experiment. }
    \label{fig:tensile}
\end{figure}

We did a tensile test to further prove that our method is mechanically as identical as the normal 3D printing. To do that, we printed two samples; The first samples are printed with 100\% infill with both normal and streaming method, and the second samples are printed with 700\% infill. Our specimen is the standard D638 V ( shown in Figure \ref{fig:labtest}). In the tensile test, a pulling force is applied to the specimen and load vs crosshead displacement of the object is identified. This curve for the first trial is shown in Figure \ref{fig:tensile}. Based on our experiment, the 100\%infill specimen could stand a higher amount of force before the breakage point, which was 404.05 $\pm$ 11.59(N) for the normal method and 396.27(N) for the streaming method, which shows a small amount of difference (~1.9\%) from the average of normal trials (see Table \ref{table:result}). The 70\% infill specimen in the normal method broke at 393.98 $\pm$ 19.12(N) while in the streaming case the maximum load before failure was 367.35(N) (~6.5\% difference from the average in the normal case).

\begin{table}[!b]
\centering

\caption{Tensile test results}
\label{table:result}

% \begin{table}
    \centering
    \begin{tabular}{ccc}
    \toprule
    
         & Infill 100\% Normal  & Infill 70\% Normal\\
             \midrule

         Trial 1 &387.96  &409.29 \\
         Trial 2 &419.16  &364.45 \\
         Trial 3 &399.52  &389.78 \\
         Trial 4 &409.56  &412.41 \\
       Mean $\pm$ STD  & 404.05 $\pm$ 11.59 &  393.98$\pm$ 19.12 \\
\bottomrule
 
 \end{tabular}

\end{table}

% \textcolor{red}{We need to write an analysis on how this process prevents attacks or improves security.}

\section{Conclusions and Future Work}
\label{sec:conclusion}
    In this work, we introduced an STL streaming framework the aims to protect IP and enhance security in a MaaS platform based on 3D printing. By streaming STL files layer by layer to the manufacturer, the risk of IP theft is significantly reduced. Our framework uses the custom python scripts to segment the original STL file and convert it into STL sections. The proposed method demonstrated the effectiveness in printing both simple designs as well as those requiring special orientations or additional support. Compared to previous works that stream G-code files that can be infected and thereby damage the manufacturing process and machine, our method is robust to these attack scenarios and manufacturers can safely generate G-code files on their side. 
    A limitation that can be addressed as a future effort is that there is no guarantee that the manufacturer would delete the previous slices and not save all of them together. In that case, after the completion of a streaming process, a malicious manufacturer can put all these slices together and reconstruct the design file. To solve this issue, it is necessary for the manufacturing machines to have a streaming-like application that prohibits workers from accessing the local memory.


% \begin{acks}
% This work was supported by the NSF Grant CCRI 2234972. Any opinions, findings, conclusions, or recommendations expressed in this material do not necessarily reflect the views of the funding organizations.
% \end{acks}
\begin{acks}
This work was supported by the NSF Grant CCRI 2234972. Any opinions, findings, conclusions, or recommendations expressed in this material do not necessarily reflect the views of the funding organizations.
\end{acks}

% \newpage
% \newpage

% \bibliographystyle{ACM-Reference-Format}
% \bibliography{refs}


\end{document}